\documentclass[preprint]{aastex62}
\usepackage{xcolor}
\usepackage{multirow}
\usepackage[normalem]{ulem}
\usepackage{physics}
\usepackage{cleveref}
\usepackage{graphicx}
\usepackage{epstopdf}
\usepackage{natbib}

\accepted{\today}
\submitjournal{ApJ}

\begin{document}

\title{On Solar Recurrent Coronal Jets: \\ Coronal Geysers as Sources of Electron Beams and Interplanetary Type-III Radio Bursts}

\correspondingauthor{Alin Razvan Paraschiv}
\email{alin.paraschiv at monash.edu; paraschiv.alinrazvan at gmail.com}

\author[0000-0002-3491-1983]{Alin Razvan Paraschiv}
\affiliation{School of Mathematical Sciences\\
Monash University \\
9 Rainforest Walk \\
Clayton, Victoria 3800, Australia}

\author{Alina Donea}
\affiliation{School of Mathematical Sciences\\
Monash University \\
9 Rainforest Walk \\
Clayton, Victoria 3800, Australia}

\begin{abstract}
Coronal Jets are transitory small-scale eruptions omnipresent in solar observations. Active regions jets produce significant perturbations on the ambient solar atmosphere and are believed to be generated by microflare reconnection. 

Multiple sets of recurrent jets are identified in extreme-ultraviolet filter imaging. In this work we analyze the long time-scale recurrence of  coronal jets originating from a unique footpoint structure observed in the lower corona.  We report the detection of penumbral magnetic structures in the lower corona. These structures, that we entitled "Coronal Geysers", persist through multiple reconnection events that trigger recurrent jets in a quasi-periodical trend. 

Recurrent jet eruptions have been associated with Type-III radio bursts that are manifestations of traveling non-thermal electron beams.We examine the assumed link, as the coronal sources of interplanetary Type-III bursts are still open for debate. We scrutinized the hypothesized association by temporally correlating a statistically significant sample of six Geyser structures, that released at least 50 recurrent jets, with correspondent Type-III radio bursts detected in the interplanetary medium. 

Data analysis of these phenomena provides new information towards understanding of small-scale reconnection, non-thermal electron beam acceleration, and energy release. We find that the penumbral Geyser-like flaring structures produce recurring jets. They can be  long-lived, quasi-stable, and act as coronal sources for Type-III bursts and implicitly for upwards accelerated electron beams. 
\end{abstract}


\section{Introduction}

Coronal jets linked to active regions (AR) recently became a very active research topic due to the improvement in imaging and cadence of the available instrumentation. The magnetic topology associated with ARs tends to usually be more complex, making AR coronal jets hotter and larger when compared to polar jets \citep{moore2010,sako2013}. Current day studies do not necessarily restrict possible `jets' to the complete `traditional' definition as the term became loosely interpreted. For the purpose of our study, we hypothesize that jets are associated with low-lying local microflaring footpoints and emphasize their escape into the inner heliosphere. \citet{shimojo1996,shimojo2000} studied X-ray jets originating in active regions, quiet sun, and coronal holes, finding that $\sim$68\% of the jets appear in or near ARs,  proposing an associations between coronal jets and micro or nano class flares.  

Short time recurrent AR jet emission has been discussed in the past (e.g. \citet{guo2013}, $\sim$1 h temporal interval; \citet{innes2011}, $\sim$2 h temporal interval; \citet{chifor2008}, $\sim$3 h temporal interval; etc.). Observations of recurring jet were also discussed by \citet{chifor2008} using images from the Extreme Ultraviolet Imagers \citep[EUVI;][]{wuelser2004} instruments of the Solar Terrestrial Relations Observatory \citep[STEREO;][]{kaiser2008}. They associated recurring magnetic flux cancellations close to a pore with corresponding X-ray jet emissions and chromospheric ribbon brightenings. The observations from the Solar Dynamics Observatory, \citep[SDO;][]{pesnell2012} have proved essential to the advancement of coronal jet physics. Using observations in Extreme UltraViolet (EUV) from the Atmospheric Imaging Assembly \citep[AIA;][]{lemen2012} and data from the Helioseismic and Magnetic Imager \citep[HMI;][]{scherrer2012}, \citet{guo2013} and \citet{schmieder2013} analyzed three short-time recurring AR jets aiming to understand their morphology, and reported twisting motions.  Recurring jets have been simulated. Multiple competing scenarios were proposed \citep[e.g.][]{moreno2013,archontis2013,pariat2015,cheung2015,lee2015} and particular recurrent jet observations have been reproduced. A generalized model depiction of recurrent jets is not yet attainable as the interpretations are still debated, see \citet[chap. 9]{raouafi2016}.   

EUV and X-ray jet eruptions with detectable microflaring activity at their footpoints were associated with co-temporal Type-III radio bursts \citep{kundu1995}. Solar radio bursts are classified based on radio dynamic spectra morphology and drift rate, and sometimes are associated with solar eruptive activity. Type-III radio bursts are a subclass representing fast drifting enhancements in the radio dynamic spectra which are interpreted as a signature of non-thermal electron beams that escape into the interplanetary medium. \citet{krucker2011} suggested that non-thermal electron events are released by magnetic reconnection events between open and closed magnetic field lines in a spire configuration. \citet{innes2011} showed that three recurrent jets originating in an AR penumbra qualitatively correlate well with detected Type-III radio bursts. This has been extended by \citet{mulay2016}, who reported correlations of different AR jets with radio bursts.  

The case studies of \citet{mccauley2017} and \citet{cairns2018} use ground-based radio observations aiming to probe the reconnection processes that generate coronal jets and the observed Type-III bursts. A complementary study is presented by \citet{mccauley2018}, who derived density determinations for Type-III bursts that were associated to coronal streamers. In this case, the authors utilized observationally driven white-light streamer coronal density models, discussed electron beam propagation effects that may influence the source spatial position, but did not clarify the location of the EUV sources for their sample of three analyzed bursts.

A comprehensive review on Type-III emission is presented in \citet{reid2014}. Currently, Type-III burst generation is linked to flaring (or microflaring) events in the lower corona \citep[see][chap. 2.5.2]{reid2014}, and is associated with either CME's or jets in a number of case studies \citep[see][chap. 7.1]{raouafi2016}. Aspects such as broader definitive correlation between coronal sources and Type-III bursts, electron beam production from microflare sites, and jet vs. CME source type are still elusive and worthy of further discussion. We aim to push this discussion further, by proposing that long lived, quasi-stable, EUV and X-Ray jet reconnection sites, labeled herein as coronal Geysers, are `trustworthy' coronal sources for Type-III bursts. 

We aim to clarify the nature and source of Type-III burst and provide a physical interpretation of their association with coronal jets. Section \ref{chap:beam:secobs} covers a comprehensive selection of Geyser datasets, utilized instrumentation, and data interpretation. Section \ref{chap:beam:sec_ts} presents a thorough analysis of detected temporal delay between EUV emission and interplanetary Type-III bursts. The observational results are cross-checked using an analytical calculation of electron beam travel time by assuming a heliospheric density model. A good agreement was found between the two independent approaches. Section \ref{chap:summ} presents a summary of the main findings, discusses the results in the context of current coronal jets and Type-III association and ends with a brief conclusion and future prospects.  

\section{Observations, Instruments, Datasets, and Interpretation}\label{chap:beam:secobs}
\subsection{Standard flares and particle acceleration}
\begin{figure*}[!ht]
  \begin{center}
	\includegraphics[width=0.92\linewidth,trim={0cm 0.33cm 0cm 0cm},clip]{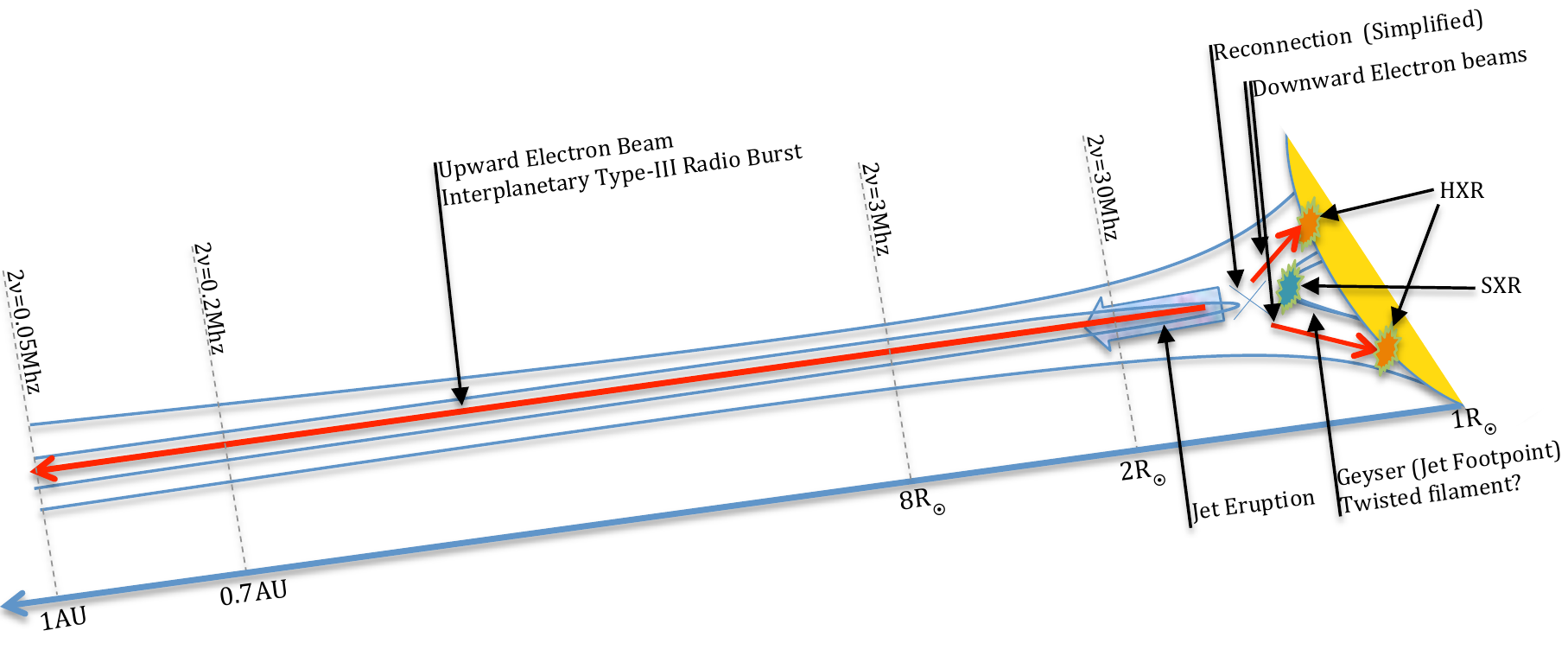}
  \end{center}
\vspace{-0.5cm} \caption{Cartoon model of coronal jets. Qualitative schematic of jet eruption features is presented along with the associated upwards and downwards electron beams. The upwards beam propagation into the heliosphere is drawn based on the observational properties of generated Type-III radio bursts and is compatible to coronal density models \citep[e.g.][]{mann1999}. The cartoon is consistent with the widely accepted CSHKP flare scenario.}
\label{chap:beam:figcartoon1}
\end{figure*}

We describe the small-scale jet eruptions using the standard flare frameset \citep[CSHKP; see][]{carmichael1964,sturrock1966,hirayama1974,kopp1976} including particle acceleration mechanisms \citep[see review; ][]{shibata2011}. We assume that particle acceleration mechanisms are a product of the magnetic reconnection processes. \citet{shibata2011} also predicts the existence of downwards beams. These stream towards the flare footpoints along the newly reconnected field lines. In a complementary study \citep{paraschiv2018}, we have observed RHESSI \citep{lin2002} hard X-ray emission in the 12-25 Kev energy range during the Geyser's successive flaring events. This suggests the existence of high energy tails in the electron distribution, which are consistent with beams. The most accepted model for X-ray emission is the thick-target bremsstrahlung. We do not address herein the downwards particle acceleration or X-ray emission.  The \cref{chap:beam:figcartoon1} sketch presents our generalized schematic view of a small-scale reconnection-driven jet, where upwards and downwards beams are generated, strengthening the assumption that small-scale impulsive (micro)flaring processes are involved in jet production. To this extent, the  observational features are presented under the assumption that the magnetic reconnection is involved in generating unstable electron beams. This may not always be the case. 

\subsection{The EUV coronal jets of the AR11302 Geyser site}

We define \textbf{Coronal Geysers }as long-lived small-scale penumbral active region structures that have an open field coronal connectivity, are prolific generators of recurrent jet eruptions, sources of particle acceleration and radio bursts, an are classified from  an energetic point of view as impulsive microflare sites. The geyser structures have roots in complex magnetic topologies, are subject to helicity conservation, and can contain filamentary structures. In this work, we focus on the upwards non-thermal particle beam that are generated simultaneously with coronal jets.

A small-scale persistent recurrent jet site, labeled G1, was observed at the south-eastern periphery of AR11302 on 25 Sep. 2011.  The AR was prolific in generating a variety of  eruptive events: X-class and M-class flares, complex CME eruptions, small-scale events, etc. By small-scale we understand eruptions (flares) which are roughly at least one order of magnitude smaller in scale and power when compared to the standard ($\ge$B class) flares. A typical microflare event dissipates energies in the order of $\le 10^{-5}$ erg cm$^{-2}$ s$^{-1}$, with total energies $<10^{27}$ erg or $<10^6$ W m$^{-2}$, as observed in 1-8 \AA{} X-ray flux measurements \citep{hannah2011}.    

We use multiwavelength observations in the EUV channels of SDO-AIA. The AIA EUV imager provides the highest resolution full-disk solar images, observing the sun in ten UV and EUV channels, with a spatial resolution of $\sim 0.6'' \cdot$ pix$^{-1}$ and a temporal cadence of 12 s. The raw data were extracted via the JSOC pipeline and calibrated locally to level 1.5. The data was further processed using the Solarsoft (SSWIDL) package and additional corrections were applied, e.g. pointing, co-alignment, respiking, aia\_prep corrections, etc. 

The jet activity is linked to the host active region by \citet{donea2013} and \citet{cairns2018}. We detected numerous EUV jets associated with an unique footpoint, the G1 Geyser, during multiple days of AR11302's near-side visible lifetime.  We selected a 24 hours timespan, covering the date of 25 Sep. 2011, since this day was most prolific in generating jets.  We selected a 13$''$ x 17$''$ region centered around the G1 footpoint and applied an inhouse developed full cadence spatial tracking procedure to the AIA-171\AA{}  and AIA-304\AA{} data. The procedure computes a timeseries of the total or pixel averaged intensity inside a selected region while accounting for solar projection and rotation effects that manifest when dealing with a long temporal tracking. The resulting intensity light-curves are presented in the top panel of \cref{chap:beam:fig-correl}. 

Ten individual jets, J1 to J10, were detected as originating from the G1 structure during our 24 h observation. The G1:J1 - G1:J6, and G1:J8 EUV jets can clearly be identified, whereas the events G1:J7, G1:J9, and G1:J10  have shorter lifetimes and smaller intensity flaring. \citet{cairns2018} based their work on the radio observations of the  Murchison Widefield Array \citep[MWA;][]{tingay2013}, analyzing radio manifestations of `jet' activity of the same G1 footpoint for a much shorter time interval, between 01:11UT and 01:24UT, 25 Sep. 2011. In our EUV recurrent jet timeseries  the flaring episode corresponds to two distinct jets, labeled as G1:J2 and G1:J3 that are occurring during the subinterval. We also observed minor flaring events, spatially correlated with the Geyser locations. These events manifested individually or in groups, but were not accompanied by jet emission in any of the SDO-AIA channels. From the EUV perspective, all G1 jets followed the same propagation direction, erupting along an apparently `open' magnetic structure. 

\begin{figure}[!ht]
\begin{center}
	\includegraphics[width=0.49\linewidth]{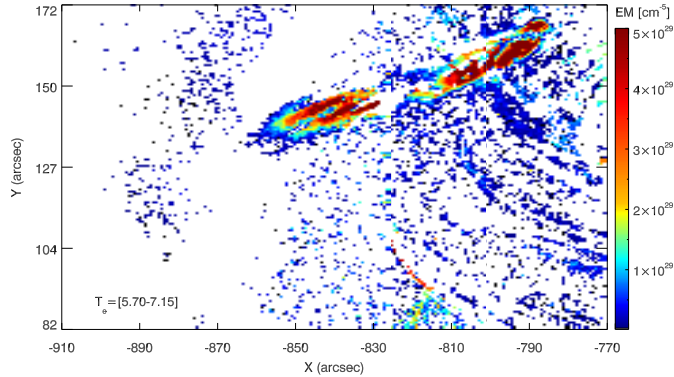}
  \end{center}
\vspace{-0.5cm} \caption{Observational portrayal of coronal jets. The differential emission map portraying the strands comprising one coronal jet, G1:J6 13:08UT of the AR11302 Geyser site. Differential emission measure is recovered via the \citet{hannah2012} regularized inversion. }
\label{chap:beam:figcartoon2}
\end{figure}

We briefly describe one jet eruption (G1:J6) with respect to the emission measure recorded by the SDO-AIA filters. We recovered the Differential Emission Measure (DEM) peak emitting temperatures using the regularized inversion method developed by \citet{hannah2012} applied to the SDO-AIA filter set. The DEM recovered observational morphology of the G1:J6 jet is presented in \cref{chap:beam:figcartoon2}. It is worth mentioning that this is a particularly impulsive jet. The analysis revealed two distinct temperature peaks: one at $\log T_e\sim 6.3\pm$0.1 K ($\sim 2$ MK) corresponding to the main jet eruption; the other peak is more pronounced and associated to the Geyser hot loops, centered around $\log T_e\sim6.9\pm$0.12 K($\sim 8$ MK) . The lower temperature and its correspondent electron density estimation, $n_e\sim0.5-1\cdot 10^{10}$ cm$^{-3}$, is consistent with the interpretation provided by \citet{mulay2016}.  This density limit places the G1 Geyser  at a height of $\sim$2-3 Mm above the photosphere according to general heliospheric density models \citep{mann1999, fontenla2011}. These parameters are used as boundary limits for the heliospheric density model discussed in Sec. \ref{chap:beam:sec_ts:subsecc_model}. A detailed DEM analysis, along with interpretation of X-Ray emission, and the downwards particle beams associated to the G1 Geyser site are treated separately (see Chap. 3 \citet[][PhD Thesis]{paraschiv2018}, in preparation). 

\subsection{EUV coronal jets from other AR Geysers}

To complement our work, we searched the available literature and selected five additional recurrent jet sites that manifested similarly to G1. The sites produced coronal EUV jets that were studied by \citet{sterling2016,chen2015,hu2016,panesar2016}, and \citet{liu2016}, which we incorporated in our analysis. Importantly, these studies do not address the association between EUV eruptions and radio bursts. A summary of observation parameters and references for all six sites can be found in \Cref{chap:beam:table-dataset}. These additional sites have similar sizes to G1, with the exception of the G5 site (34$''$ x 30$''$) which is also associated to more energetic flaring. We chose to limit the temporal tracking to include only the jets discussed in each cited reference. Similarly to G1, jets originating from the G2-G6 sites could be detected outside the selected tracking limits.  Using the same procedures as for our G1 site, we tracked the G2, G3, G4, and G6 Geysers over time intervals between 7 h and 11 h. The G5 dataset tracking is performed over a 22 h interval, similar to the G1 site.

Geyser structures G2 and G5 are linked \citep{sterling2016,panesar2016} in part to flares of magnitudes $B$ and sometimes $C$, although most eruptive events are classified as microflares. The other sites, namely G1, G3, G4, and G6, are identified as typical microflare sites, with only very few events exhibiting stronger eruptions. 

The dataset selection is not exhaustive as our goal is to understand recurrent  jets that span across longer timeframes, hinting at the existence of quasi-stable Geyser structures.  A total of 40 out of 44 reported jet eruptions could be reproduced analogous to the original studies, in addition to our initial 10 G1 jets. 

\subsection{Heliospheric Type-III radio bursts}

The upward propagating Type-III radio bursts, are also ubiquitous events present in synoptic radio dynamic spectra (\cref{chap:beam:fig-wind}). They represent observational signatures of coherent electron beams that stream through `open' magnetic fields in the high solar corona and inner heliosphere \citep[see reviews; ][and references therein]{reid2014,shibata2011}. A schematic of the processes is also presented in \cref{chap:beam:figcartoon1}. The electron beams become bump-on-tail unstable \citep{sarkar2015} and interact with the local plasma via a set of non-linear processes as they pass along the medium. As a result, plasma oscillations at the local plasma frequency ($\nu_{pe}$) and its harmonic ($2\nu_{pe}$) are generated and are sensitive to the plasma electron density ($n_e$), 
\begin{equation}
\nu_{pe}=\sqrt{\frac{e^2\cdot n_e}{\pi\cdot m_e}}\, \approx \, 8978.47\cdot\sqrt{n_e}\,,\label{chap:beam:eq-nefreq}
\end{equation}
 where $n_e$ and $\nu_{pe}$  are in units of  cm$^{-3}$ and Hz, respectively. These oscillations manifest observationally as radio bursts. The upward Type-III bursts are an observational subclass of low frequency radio emission. They are characterized by a fast frequency drift in the radio spectra maps occurring due to the mildly relativistic propagation speed ($\beta=v/c\sim 0.1-0.3$) of the electron beams. See \citet{reid2014} review for a detailed description of the topic. 

 Images from the  EUVI-195\AA{} instruments onboard STEREO twin observatories show different vantage points of the jets, compared to SDO-AIA.  The STEREO EUVI-195\AA{} predominantly observed the flaring of the jets Geyser footpoint.  The main outflow jet emissions remained hardly detectable due to projection effects combined with the lower sensitivity, spatial ($ 1.1''$ pix$^{-1}$) and temporal (300 s) resolution of EUVI-195\AA{}. The WIND \citep[WAVES;][]{bougeret1995} and STEREO \citep[SWAVES;][]{bougeret2008} radio data were used to investigate the relationship between  interplanetary Type-III radio bursts and the Geyser generated jet eruptions.
 
 Assuming a coronal and inner heliospheric density distribution one can theoretically reproduce the travel of upwards electron beams propagating into the corona that are detected as Type-III radio bursts. For this, we require a starting $\nu_{pe}\approx 300-500$ MHz frequency, corresponding to a chromospheric/coronal reconnection site, and an end $\nu_{pe}\approx 0.02-0.2$ MHz frequency corresponding to 1AU distance. In practice, such a determination is limited by instrumental constraints. The STEREO and WIND instruments are covering a frequency range from $16$ MHz to $0.01$ MHz and have a temporal cadence of 60 s. The 16 MHz  instrumental limit only allows the recovery of radio emission from plasma in the interplanetary space. In \cref{chap:beam:fig-correl} the radio timeseries (blue,pink, and green curves) were interpolated in order to reproduce the SDO-AIA temporal cadence, but discussed uncertainties remain set to 1 datapoint per minute.

\section{Coronal Geyser Sites and Type-III Bursts}\label{chap:beam:sec_ts}

\subsection{Temporal correlation of EUV coronal jets and Type-III radio bursts}\label{chap:beam:sec_ts:subsecc_correl}
    
\begin{figure}[!ht]
\begin{center}
\includegraphics[width=0.85\linewidth]{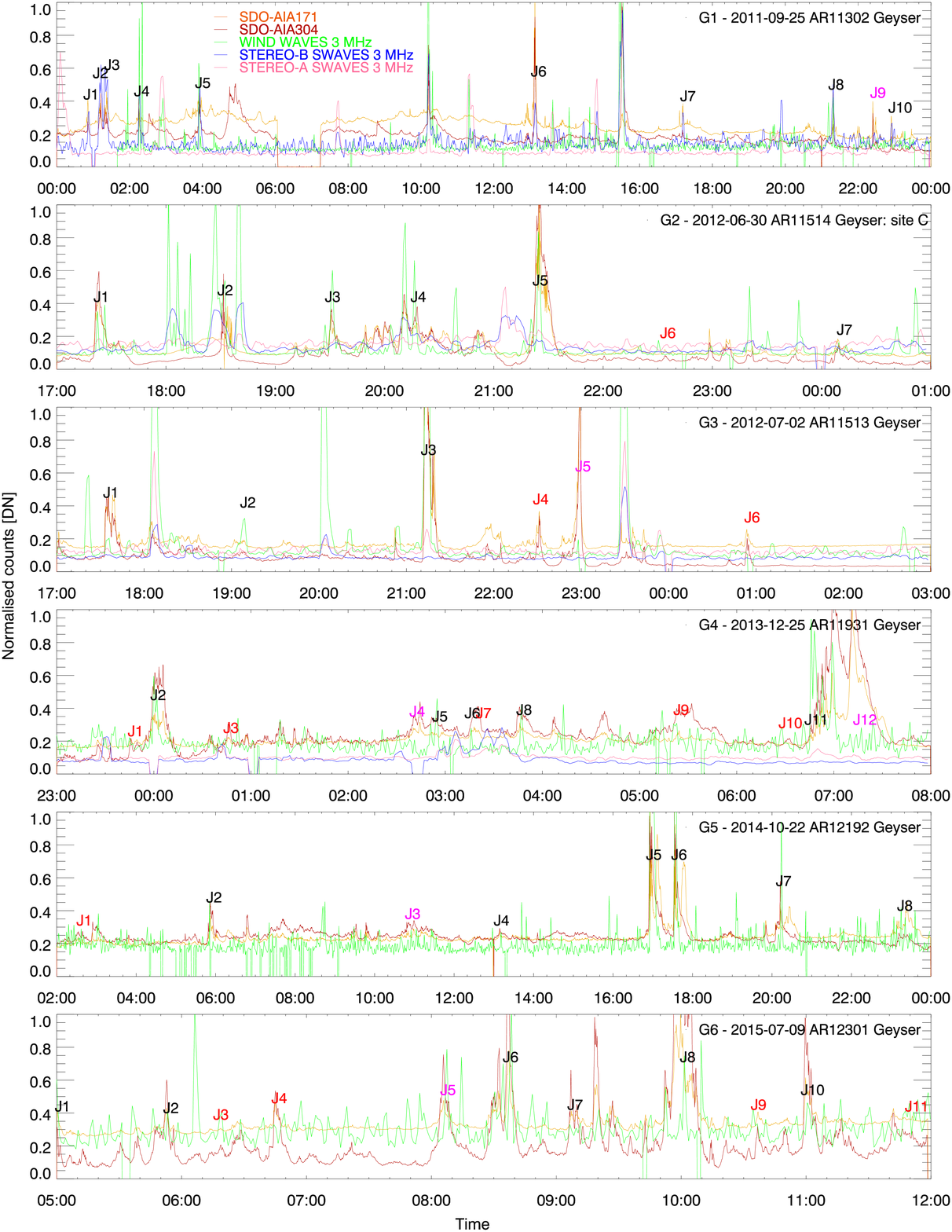} 
  \vspace{-0.5cm}\caption{Temporal tracking of the solar regions centered on the studied Geyser structures using SDO's AIA-171\AA (orange), AIA-304\AA (red), and 3 MHz channel data from STEREO-B SWAVES (blue), STEREO-A SWAVES (pink), and WIND WAVES (green). The individual jets are labeled incrementaly for each referenced dataset. The x-axis (time) shows different temporal intervals, as correspondent to each dataset.  Uncorrelated events (redlabels) and uncertain events (purple labels) are highlighted. The radio data is slightly shifted along the y-axis to increase plot readability. Additional information on the six datasets can be found in \Cref{chap:beam:table-dataset}.}\label{chap:beam:fig-correl}
\end{center}
\end{figure}    
    
To verify the assumptions of a direct correlation between the EUV jets and interplanetary Type-III radio bursts we superposed the EUV Geyser timeseries with the STEREO SWAVES and WIND WAVES dynamic spectrum radio data for all discussed sites and jet eruptions. The Geyser sites were usually observed by multiple satellites, each providing a unique vantage point for radio burst interpretation. The recorded STEREO and WIND radio signal is integrated over the entire solar disk. SDO and WIND share co-spatial viewpoints, while the two STEREO satellites have uniquely different viewpoints. This increases the possibility of positive correlations by superposing multiple sources that observe the site. Negative correlations are possible when particular satellites do not observe the flaring site. 
 
All G1-G6 sites were observed in EUV by SDO-AIA. The WIND signal corresponds to the same solar disk viewpoint. The G2, G4, G5, and G6 sites were observed close to a disk central location. G1 and G3 were observed at higher longitude positions. The G1 site is located on the eastern side of AR11302. Due to the separation between the SDO and STEREO-B spacecrafts ($\sim 97^\circ$), the AR also appears on the western side of STEREO-B's visible disk. Events were positively correlated for both the WIND and STEREO-B data, although few jets were missed by WIND due to datagaps. The site was not visible from the STEREO-A viewpoint. No coincidental association was found. In other words, all 10 events were negatively correlated using the STEREO-A viewpoint. 
 
 Geyser site G3  and AR11513 were situated at the western side of the solar disk. Due to the separation  between the SDO and STEREO-A spacecraft ($\sim 119^\circ$) the AR also appears on the eastern side of STEREO-A's visible disk, and the EUV jet to Type-III burst association could also be established. No coincidental association could be found using STEREO-B, which does not observe the site. We report that this negative correlation holds for all events linked to the G1-G4 datasets, and no coincidental and/or falsely associated events was possible, strengthening our assumptions regarding the positive correlations. The G5 and G6 sites have no STEREO data associated due to the STEREO satellites being out of contact during their far-side crossing.

We measured the time difference between the  peak emission in the AIA-171\AA{} and the subsequent associated radio burst in the 3 MHz channel. When that was not possible, the time  difference between the onset of the two signal peaks was used, and the respective events were marked as uncertain. There was no discernible difference when estimating the time difference via the onset or the peak emission datapoints.  Also, the AIA-304\AA{} signal is sometimes time-delayed when compared to the AIA-171\AA{} making it unsuitable for this task. For particular events, there may be multiple bursts generated corresponding to multiple SDO-AIA flaring peaks occurring in short timescales of $<150$ s (e.g. G1 site, events J2 and J3). 

In total, we have analyzed six sets of EUV recurrent jets, where 50 individual eruptions were detected. \Cref{chap:beam:fig-correl} shows the superposed timeseries data for all sites. \Cref{chap:beam:table-dataset} presents the observed EUV to radio positive and negative correlations. Uncertainties for all Geyser datasets are included.   We have found that 35 EUV jets plus 6 eruptions of weaker correlation were positively associated to propagating Type-III interplanetary radio bursts. No jet eruptions could be negatively associated to close-following ($<180$ s) Type-III radio emission, negating the possibility of accidental associations.  As the possibility of accidental associations is significant, the 0 event negative correlation is important.  

A significant set of underlying assumptions and uncertainties involved are considered: 
The temporal correlations are influenced by the low data cadence of the radio signal. Additionally, radio data may be of low quality as the (3 MHz) signal is sometimes saturated by multiple overlapping events. To this extent, the correlations were always performed by identifying a radio burst that immediately follows the main SDO-AIA EUV site flaring, as a $\nu_{pe}$= 3 MHz emitting plasma corresponds in general to a high heliospheric altitude (see \cref{chap:beam:figcartoon1}). The radio burst signal may be weak when compared to the background, thus it is checked individually for statistical relevance by evaluating the Signal to Noise Ratio (SNR). 31 correlated events had a detection limit of $>3\sigma$ and an additional 10 events had SNR $>2\sigma$. No correlated event had a radio peak of $<2\sigma$ in the 3 MHz flux.  

The EUV Jets for sites G2-G6 were extracted based on the identifications provided by their references (\Cref{chap:beam:table-dataset}, last column). four reported jets linked to the G4 and G6 Geyser sites could not be separated or differentiated in our SDO-AIA timeseries or by visual inspection. We have not included these in the total. Jets that were not shortly followed by a radio bursts were deemed uncorrelated and labeled in red in \cref{chap:beam:fig-correl}. Purple colored events were considered weakly correlated, due to either uncertain radio burst SNR ($<3\sigma$) or unreliable time delay between the SDO-AIA and burst peak times ($\tau_{obs}<30$ s or $\tau_{obs}>120$ s). The plotted 3 MHz radio flux was linearly interpolated with intermediary 12 s datapoints to increase plot readability.

The \cref{chap:beam:fig-err} (left panel) histogram shows a normal distribution of measured time delay ($\tau_{obs}$) between the SDO-AIA flaring and the radio Type-III burst detection times. The time delay is {\em always positive}, meaning that the radio burst is subsequent to the EUV emission ($\tau_{obs}=\tau_{EUV}-\tau_{burst}>0$), as expected when assuming that the $2\nu_{pe}=3$ MHz burst signal corresponds to considerable heights. We stress that $\tau_{obs}$ requires a cautious interpretation as the cadence for SDO-AIA fluxes is 12 s while the radio data cadence is 60 s. 

Hypothetically, the 6 uncertainly correlated events can be removed from the histogram distribution. These events are located at the edge of bins, where the assumptions are less reliable, e.g. 12-24 s delays would imply a much higher burst propagation speed, and 122-144 s delays would imply a very weak electron beam (burst) acceleration.  A Gaussian fit over the time delay distribution histogram revealed an  average time delay $ \tau_{obs}=72$ s  with a standard deviation  $\sigma_{\tau_{obs}}$=22 s. 

We decided to use the full distribution (including the six uncertain jets) as we considered all 41 jets/bursts to be correlated in our analysis.  The average time delay  $ \tau_{obs}=72$ s is unchanged and is constrained by a wider $\sigma_{\tau_{obs}}=27$ s width. Additionally, the histogram has a bin size of 12 s that corresponds to the SDO-AIA cadence. We remind the reader that $ \tau_{obs}$ is close to the radio signal detection limits of 60 s discussed above, hindering the use of 60 s histograms bins as required by a rigorous uncertainty estimation. Taking into account this maximum uncertainty, we would obtain $\tau_{obs}=72\pm57$ s. Following this limitation, we chose to present the $\tau_{obs}$ estimation in terms of the 12 s bins, following the SDO-AIA cadence, resulting in an additional $\pm6$ s cadence uncertainty.  Thus, to account for both the binning and variance uncertainties we set an average time delay $\tau_{obs}=72\pm33$ s, acknowledging that the uncertainty is not fully constrained.  

\begin{table*}[!t]
\centering\footnotesize
\caption{Datasets of "geyser-like" sites that generated recurrent jets. The jet eruptions were correlated with interplanetary Type-III radio bursts. The radio data sources are STEREO-A (S-A), STEREO-B (S-B), and Wind (W).  The three instruments had unique spatial positions for each dataset observations, enabling both positive and negative correlations to be discussed. 41 out of 50 detected jets were positively correlated. No event was negatively correlated. All correlated events are checked for statistical relevance against background radio counts. }\label{chap:beam:table-dataset}
\begin{tabular}{|c|c|ll|c|c|}
  \hline
  Geyser Dataset   & Identified &  \multicolumn{2}{c|}{~~~~Type-III Radio Burst Correlation~~~~}   &  ~~~~~~Correlated~~~~~~  & \multirow{2}{*}{ Study Reference } \\
 
 Observation  &    Jets           &  ~~~~Positive     &    ~~~~~Negative.     &   Events SNR         &    \\
  \hline
G1 AR11302   &   \multirow{2}{*}{  10 }  &  ~S-B: 9+(1)/10$^B$    & \multirow{2}{*}{  S-A: 0/10 }  & ~~~~~~~~$9/10>3\sigma$      &\multirow{2}{*}{\scriptsize{  This work  }}  \\
\scriptsize{25.09.2011T00:00-23:59} &  &~W: 5+(1)/7$^B$  & &$2\sigma<1/10<3\sigma$ & \\
\hline

G2 AR11514   &  7 & \multirow{2}{*}{  W: 6/7  }   & \multirow{2}{*}{ S-B \& S-A: 0/7 } & ~~~~~~~~$5/7>3\sigma$     &\multirow{2}{*}{\scriptsize{\citet{sterling2016} }} \\
\scriptsize{30.06.2012T17:00-01:00} &\scriptsize{site C} & & &$2\sigma<1/7<3\sigma$ & \\ 
\hline

G3 AR11513  &  \multirow{2}{*}{6} & ~W: 3+(1)/6$^B$ & \multirow{2}{*}{ S-B: 0/6 }   & \multirow{2}{*}{~~~~~~~~$4/6>3\sigma$} &\multirow{2}{*}{ \scriptsize{\citet{chen2015}}} \\
\scriptsize{02.07.2012T17:00-03:00} & & ~S-A: 3/6 & & & \\ 
\hline

G4 AR11931    & \multirow{2}{*}{10/12$^A$ } & \multirow{2}{*}{ W: 5+(2)/10$^B$}&  \multirow{2}{*}{ S-B \& S-A: 0/10 }  & ~~~~~~~~$3/10>3\sigma$  &\multirow{2}{*}{\scriptsize{\citet{hu2016}}} \\
\scriptsize{25.12.2013T23:00-08:00} & & & &$2\sigma<4/10<3\sigma$ & \\ 
\hline

G5 AR12192    & \multirow{2}{*}{ 8 }  &\multirow{2}{*}{ W: 6+(1)/8$^B$ } &\multirow{2}{*}{ No STEREO data} & ~~~~~~~~$5/8>3\sigma$   &\multirow{2}{*}{\scriptsize{\citet{panesar2016}}}\\ 
\scriptsize{22.10.2014T02:00-00:00} & & & &$2\sigma<2/8<3\sigma$ & \\ 
\hline

G6 AR12301   &\multirow{2}{*}{ 9/11$^A$ } &\multirow{2}{*}{ W: 6+(1)/9$^B$} &\multirow{2}{*}{ No STEREO data} &  ~~~~~~~~$5/9>3\sigma$             & \multirow{2}{*}{\scriptsize{\citet{liu2016}}} \\
\scriptsize{09.07.2015T05:00-12:00}  & & & &$2\sigma<2/9<3\sigma$ & \\ 
\hline
\multicolumn{6}{l}{$^A$Not all reported jets described in the G4 and G6 source studies could be reproduced  in our SDO-AIA lightcurves.}\\
\multicolumn{6}{l}{$^B$ Events with uncertain temporal correlations are listed inside round brackets.}
\end{tabular}
\end{table*}

\subsection{Modeling the heliospheric travel of upward electron beams }\label{chap:beam:sec_ts:subsecc_model}

\begin{figure}[!ht]
\begin{center}
\includegraphics[width=0.85\linewidth]{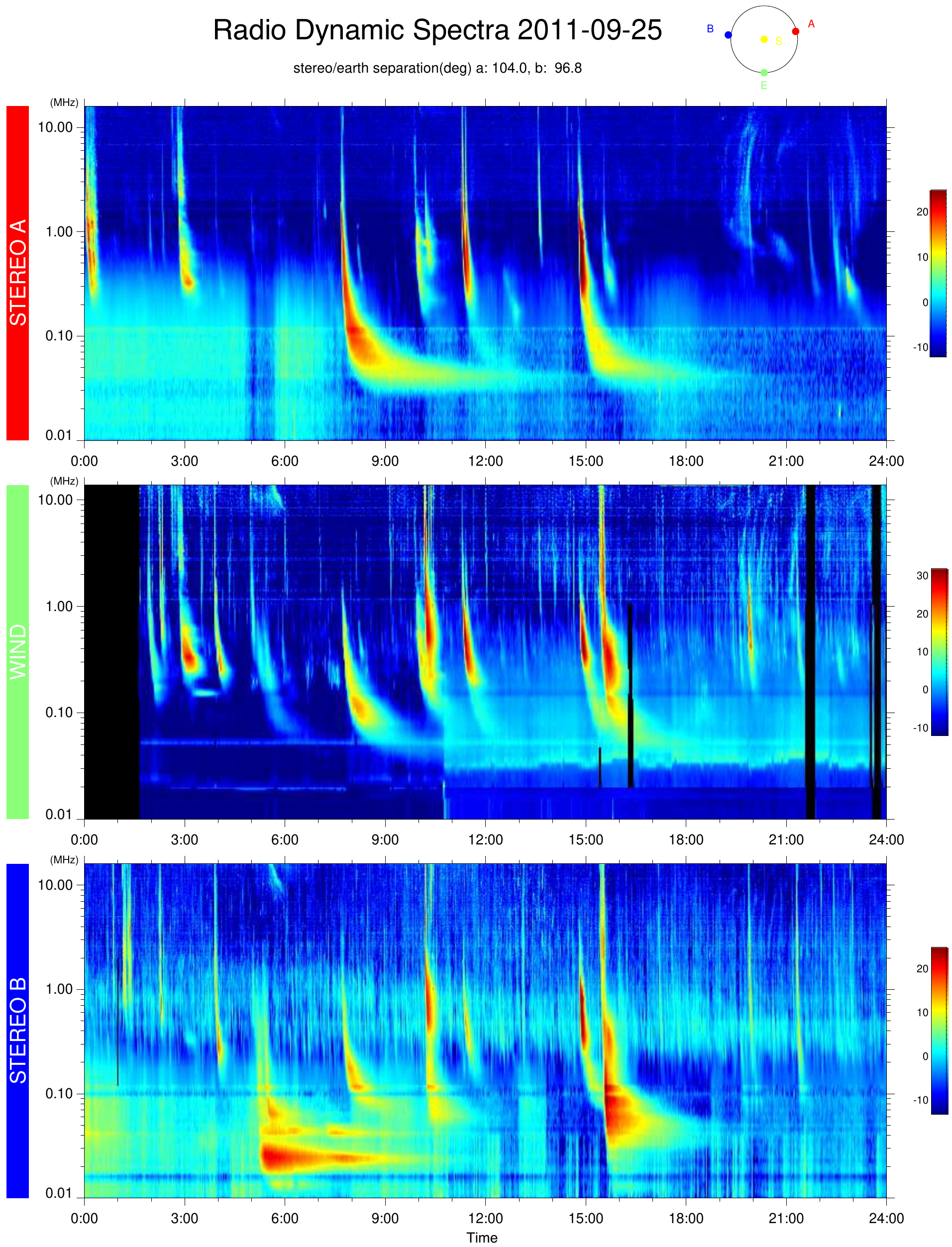}  
\vspace{-0.5cm}\caption{Dynamic spectra of the solar radio emission corresponding to the G1 site as viewed from the STEREO-B SWAVES, STEREO-A SWAVES, and WIND WAVES perspectives. The relative positions of the sattelites are shown in the upper right corner. The E (earth) position corresponds to the position of WIND and SDO.  The 11302AR appears on the western side of STEREO-B's visible disk and was detected by its SWAVES instrument. The EUV jets were temporally positively correlated with both the WIND and STEREO-B data. The site was not visible from the STEREO-A viewpoint, thus a negative correlation holds. The plot is adapted from the NASA-GSFC SWAVES service.}\label{chap:beam:fig-wind}
\end{center}
\end{figure}

The observational correlation between the thermal EUV jet emission and non-thermal Type-III radio bursts supports our initial hypothesis, although it is not a sufficing argument by itself. We further develop a simple theoretical interpretation in support of our data analysis. We assume that the burst plasma frequency is correspondent to harmonic emission following \citet{morosan2014} and \citet{reid2014}. In \cref{chap:beam:fig-correl} all three radio data sources (STEREO-A, STEREO-B, and WIND) depict the radio emission matching an oscillating plasma frequency harmonic of $2\nu_{pe}=3$ MHz, corresponding to an approximate heliospheric height, $R\sim 8.10 R_{\odot}$, where $R_{\odot}$ is the solar radius. The 3 MHz lower limit is an optimal choice as lower frequencies are more polluted by unrelated emission or higher  background noise (see, \cref{chap:beam:fig-wind} dynamic spectra). Also, the correspondent heliospheric height is smaller than $\sim$10 $R_{\odot}$ where the solar wind becomes supersonic.  

 We ask the following: Does the observed time delay $\tau_{obs}$ reflect a physical property pertaining to the correlation or is it just the result of a stochastic process? The analytical approximation of \citet{mann1999} was used to answer this question. The approximation describes an heliospheric radial density model that can be used to characterize an electron beam (observed Type-III radio burst) traveling in the inner heliosphere via \cref{chap:beam:eq-nefreq} by converting the observed emitting plasma harmonic to a local plasma density value, which in turn corresponds to a radial upwards distance from a source surface, particularly, our chromospheric site. 
 
The \citet{mann1999} approach solves the ideal continuity, momentum, and Faraday magnetohydrodynamic equations assuming radial scalar functions for the magnetic field \textbf{B}=B(r) and plasma outflow velocity \textbf{v}=v(r). The initial conditions of the equations are constrained by our observations. The v(r) term can be found by integrating the continuity equation where the mass flux is a function of the source temperature. The \citet{Parker1958} solar wind equation is thus obtained. Then v(r)$^2$ can be substituted back in the integrated form of continuity equation, which can be solved analytically for lower heliospheric heights by assuming a non-supersonic wind speed. A barometric height formula valid for a hydrostatic subsonic regimes, 
\begin{equation}
n_e(r)=n_s\cdot e^{\frac{A}{R_\odot}\cdot (\frac{R_\odot}{r} -1) } \quad \text{or}\quad
\nu_{pe}(r)=\nu_s \cdot e^{\frac{A}{2\;R_\odot}\cdot (\frac{R_\odot}{r} -1), }
\label{chap:beam:eq_freq}
\end{equation}
 can be formulated as a function of the local plasma density $n_e(r)$, or of the harmonic plasma oscillating frequency, $\nu_{pe}(r)$. 

It is important to consider potential limitations. The \citet{mann1999} model provides reasonable observational cross-validations under the  assumption that the globally averaged electrons distribution do not depend strongly on coronal structures. The observations of \citet{Koutchmy1994} showed a difference of up to three orders of magnitude in plasma density between coronal streamers, quiet equatorial regions and polar regions at $1.3 R_\odot$. Thus, we emphasize that the radial density model offers a general picture of the outer corona and inner heliosphere only if carefully utilized inside sensible hypotheses and custom tailored boundary value assumptions. 

To this extent, we follow the \citet{mann1999} recommendations and taking into account our DEM limits, we set the constant $A/R_\odot=13.83$. We fix the boundary parameters $n_s=0.5\cdot 10^{10}$ cm$^{-3}$ and $\nu_s=644$ MHz, corresponding to the DEM approximation of the plasma density at the chromospheric source discussed in Sec. \ref{chap:beam:secobs}. The parameters describe a $\sim 1-2\cdot 10^6$ K coronal site rooted $\sim$ 2 Mm (1.01 $R_\odot$) above the photosphere.  

From the dynamic spectra presented in \cref{chap:beam:fig-wind}, the average burst drift rate between STEREO's starting frequency of 16 MHz and our 3 MHz lower limit was estimated at $\Delta t$ =1.3 pix = 78$\pm$60 s, accounting for the radio signal data cadence. The $\Delta t$ = 78$\pm$60 s drift rate corresponds to a  $7.93\cdot$10$^5 \text{ to } 0.28 \cdot$10$^5$ cm$^{-3}$ density decrease when assuming that the oscillating frequency drop $16 \text { to }3$ MHz corresponds to the harmonic (2$\nu_{pe}$). 

\begin{figure*}[!h]
\begin{center}
\includegraphics[width=0.485\linewidth]{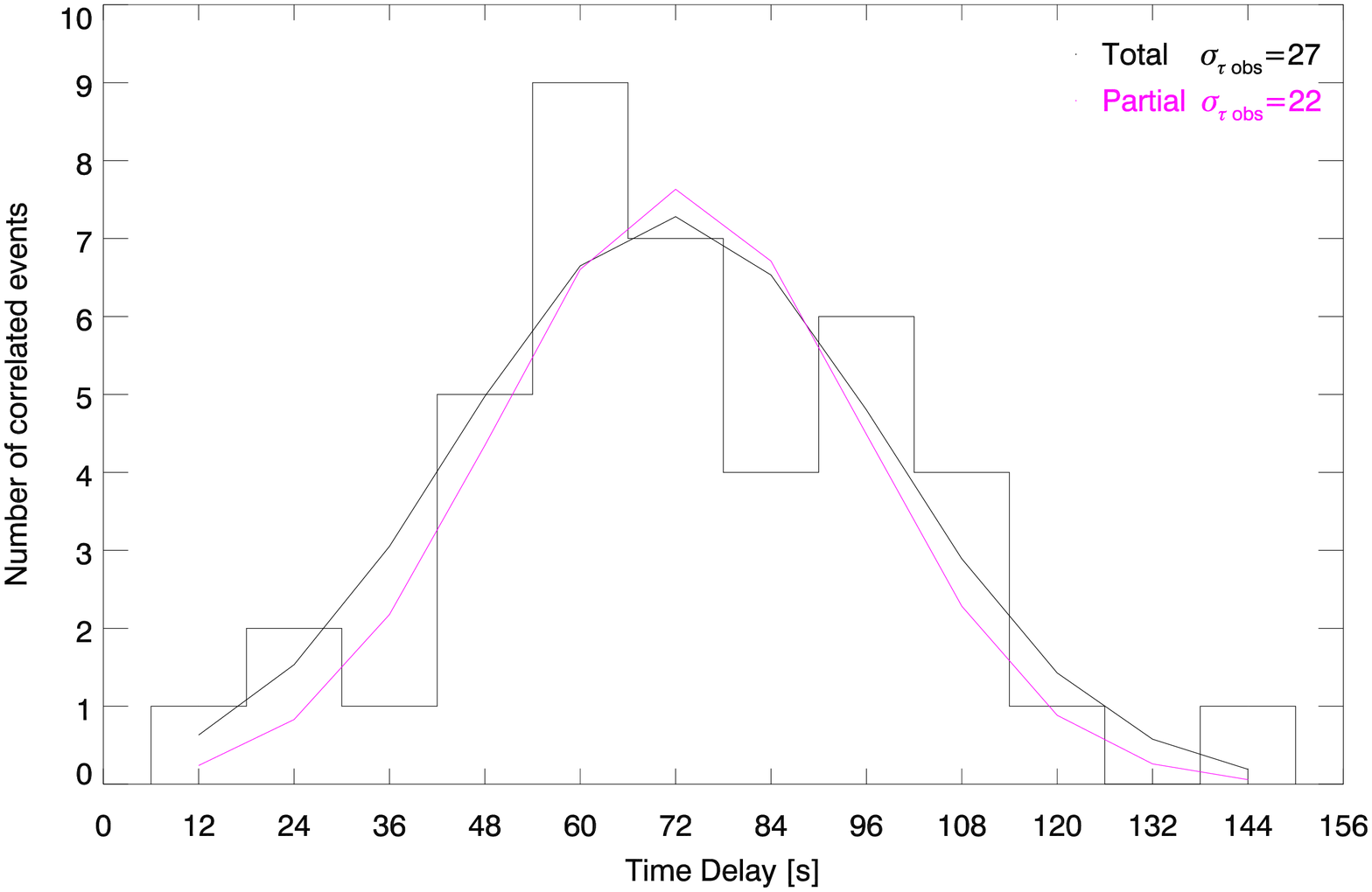}
 \includegraphics[width=0.49\linewidth]{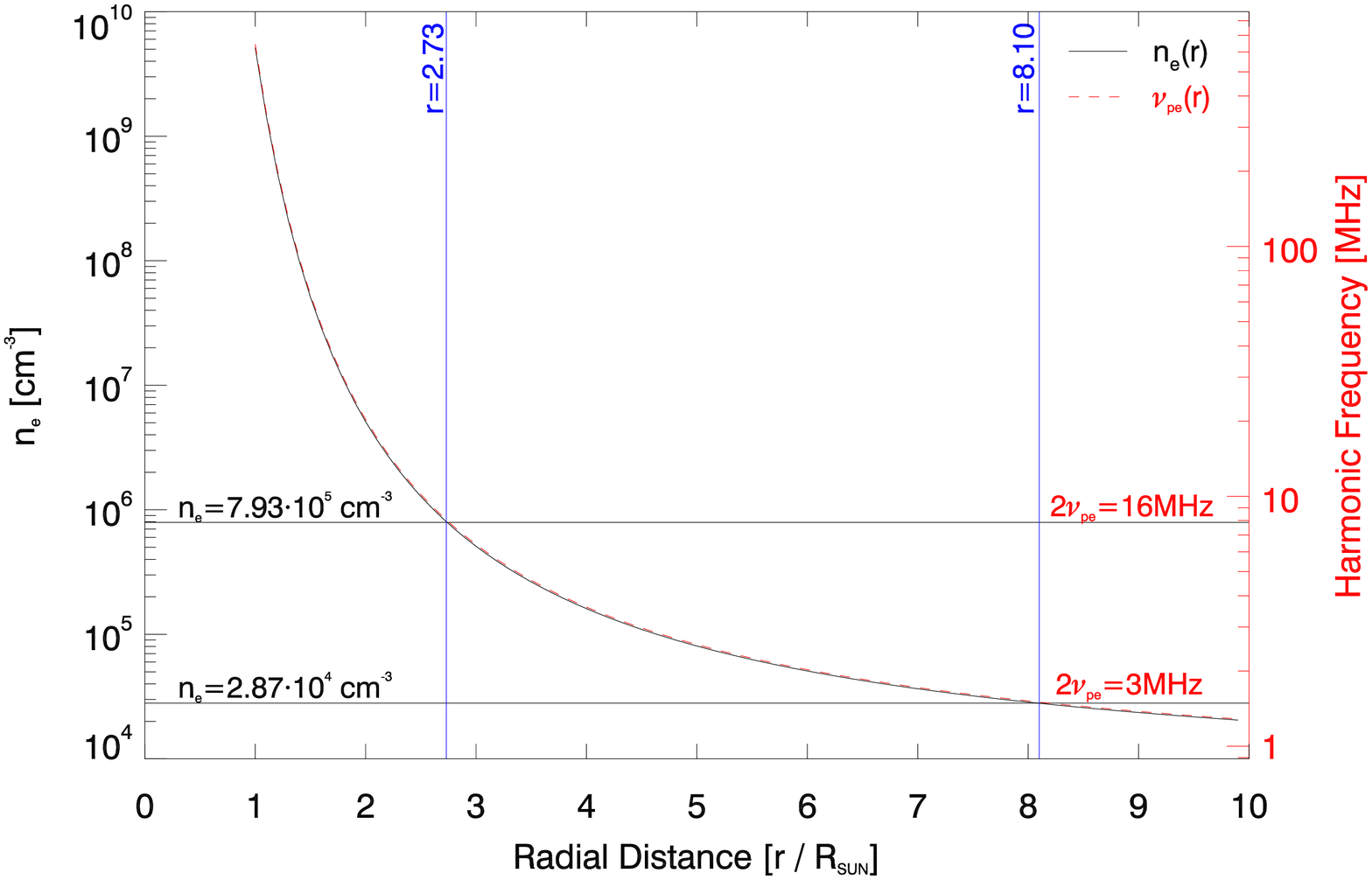}
\caption{Observational and analytical interpretation of of the recorded time delay between the SDO-AIA microflaring jet onset events and the correlated interplanetary Type-III radio bursts. \textbf{Left:}Histogram distribution of the time delay between the peak/onset SDO-AIA171\AA{} flaring and the associated Type-III bursts. The full distribution is Gaussian fitted (black curve) resulting in a average time delay  $\tau_{obs}=$72 s delay and a $\sigma_{\tau_{obs}}=27$ uncertainty. Events with $\tau\le$ 24 s and $\tau\ge 120$ s are marked as uncertain in \cref{chap:beam:fig-correl} and the distribution resulted from their omission is also fitted (purple curve). \textbf{Right:} Type-III burst drift parameters as depicted as a number density $n_e$ (black) and first harmonic frequency $2\nu_{pe}$ (red) versus the radial distance from the chromosphere via \cref{chap:beam:eq_freq}. The 16 MHz or 7.93 $\cdot 10^5$ cm$^{-3}$ signal correspond to a heliospheric height of $2.73\,R_\odot$ and the low boundary 3 MHz or 2.87 $\cdot 10^4$ cm $^{-3}$ signal corresponds to radial distance of $8.10\,R_\odot$.}\label{chap:beam:fig-err}
\end{center}
\end{figure*}

\Cref{chap:beam:fig-err} (right) shows the dependence of the heliospheric electron density $n_e$ (black) or harmonic plasma frequency $2\nu_{pe}$ (red) on the radial distance in solar radii.
The density decrease is fitted to a heliospheric travel distance of $\Delta x=8.10-2.73 =5.37\,R_\odot$, where the 8.10 $R_\odot$ point is valid when considering the subsonic approximation. This results in an averaged beam propagation speed of $v/c\sim0.16\pm0.06$ placing the detected beam in the lower end of the mild relativistic regime. 

We assume that an electron beam (observed as a radio burst) is initiated co-temporally with the SDO-AIA flaring as predicted by the standard flare model with particle acceleration. We consider the beam traveling in the $644\rightarrow 3$ MHz frequency region, corresponding to a traveled distance between $1.01\rightarrow8.10 R_\odot$. We fit the resultant beam propagation speed $v/c$ to the total traveled distance according to the density model and estimate an analytically driven time delay  between the EUV flaring and radio burst $\tau_{mod}\sim 90$ s. The estimation corroborates the observed $\tau_{obs}$=72$\pm$33 s time delay.  The $\tau_{mod}$ estimation is also influenced by the radio observations, as the $v/c$ beam speed could only be measured between $2.73\rightarrow8.10 R_\odot$.

We emphasize that a more rigorous uncertainty evaluation is not attainable as the resulting estimations are close to the instrumental sampling rates, and models have limited applicability. Modeling the travel to a longer heliospheric distance is improper as the analytic approximation holds true only for a non-supersonic solar wind. Observationally, the radio dynamic spectra cadence influences both the observed time delay $\tau_{obs}$ and the burst drift rate $\Delta t$ as both estimations are close to the 60 s data sampling. To our knowledge, no further constraints can be applied to the design of this study, as uncertainties are instrumental in nature. Radio observations at superior frequency ranges  higher cadence data sampling are required in order to increase the precision of this established correlation. 

~

\section{Summary and Discussion}\label{chap:summ}
 \subsection{Summary}
\indent\indent We identified  unique coronal sources, labeled herein as Coronal Geysers, that each produced recurrent EUV jets and multiple interplanetary Type-III radio bursts that escaped along `open' magnetic fields into the outer corona. The Geyser structures are identifiable EUV footpoints, rooted in the penumbral regions of ARs, that are subjected to recurring microflaring episodes.

Six Coronal Geysers observed by the SDO-AIA instrument are analyzed in this work. The individual temporal association between the EUV/X-ray jets and Type-III radio bursts escaping into the inner heliosphere is presented, following the standard flare model assumption that both emissions are released concurrently via magnetic reconnection. The G1 structure represents the reference dataset analyzed in this work. The G2-G6 Coronal Geysers are collected from the available literature by selecting sites that released multiple homologous/recurrent jets over a significant time period ($>$6 h), complying to our Geyser structure frame-set.  

All recurrent jet datasets, are selected with a degree of subjectivity. The G1 Geyser had a lifetime of at least one day. In our case, the tracking of G1 was limited to 24 h although the structure had a considerately longer lifetime, where jets could be observed for a period of at least 3 days. In the case of the G2-G6 Geyser structures, we limited our time intervals to describe the coronal jets analyzed in the references. We detect additional jet eruptions outside the discussed time intervals. A longer timeseries analysis on the G2-G6 Geyser sites may reveal longer lifetime scales, analogous to the G1 site. 

The utilized references (see \Cref{chap:beam:table-dataset}) discussed EUV  jets reporting similar morphological jet and footpoint properties to G1. Due to the apparent topological variations between small-scale coronal sources, we did not assess here whether the coronal Geyser frame-set can further be extended to encompass a common magnetic topological configuration. The question will be addressed in a subsequent analysis (see Chap. 4 \citet[][PhD Thesis]{paraschiv2018}, in preparation).

Not all described events are typical microflares. A subset of the discussed events, notably in the G2 and G5 sites, are associated to stronger flaring. Our correlation was not influenced by the resulting high variability in flaring power.  This indicates that, as expected, the electron beam generation is ubiquitous across the  different scales of coronal jet events.

We have identified a total of 50 jets generated by the six Geyser sites. $82\%$  (35 plus 6 uncertain jet events)  were positively associated to propagating Type-III interplanetary radio bursts. A negative correlation analysis yielded that no established correlation is false. All radio burst peaks had at least a 2$\sigma$ SNR compared to radio background.  Nine EUV events could not be correlated to Type-III radio bursts. Four jets claimed in the literature sourced could not be detected by us in the EUV timeseries flux (see \cref{chap:beam:fig-correl}). These missing jets were not counted towards the total.

We have reconstructed the time delay between the onset of SDO-AIA flaring and interplanetary Type-III radio bursts: an analytically derived time delay of $\tau_{mod}\sim$ 90 s (see \cref{chap:beam:fig-err}, right) is comparable to the observational results, where the centroid of the Gaussian fit applied to the 41 SDO-AIA vs Type-III radio bursts, revealed a average time delay $\tau_{obs}$= 72$\pm$33 s (\cref{chap:beam:fig-err}, left). Given the finite number of studied jets, we acknowledge the limitations of our statistics, and present the results as a baseline estimation. 

\subsection{EUV jets and microfilament eruptions}
 The quasi-stable Geyser structures interact via reconnection with the magnetic canopy of the ARs. Topologically they can be classified as twisted microfilaments, similar to the description provided by \citet{sterling2015}, although the dichotomy predicament is still under debate \citep[see][]{raouafi2016}. The \citet{sterling2015} microfilament eruption model assumes the existence of  multiple reconnection events that constitute an observed blowout jet eruption \citep[See also the blowout eruption cartoon of ][]{moore2010}. The observed EUV jet's threaded morphological features are compatible with an eruptive filamentary structure. The G1: J2, J3, J5 and J6 jets were highly dynamic and exhibited multiple Type-III bursts manifesting due to electron beams generated co-temporally with the multiple SDO-AIA EUV flaring peaks. This aspect represents a good indicator that multiple individual reconnection events are occurring in very short timescales. From a radio perspective, multiple short ($<$2-3 min) timed `electron injections' (electron beam generation) are reported by \citet{mccauley2017} in their high resolution MWA Type-III bursts study. Assuming the standard flare model, we hypothesize that `short electron injections' may be produced in the flaring stages of microfilament eruptions. 

 Additional evidence is provided by the \citet{cairns2018} where a discrepancy between higher frequency  (MWA, Learmonth, etc.) and low frequency (STEREO and WIND) observations is presented. The higher frequency Type-III radio bursts matched their interplanetary counterparts for the G1:J3 event. This did not occur for the G1:J2 event which does not exhibit any higher frequency signatures (see, \citet{cairns2018}, fig. 4). Our timeseries analysis (\cref{chap:beam:fig-correl}) showed that three SDO-AIA flaring peaks existed for the three interplanetary Type-III bursts associated to G1:J2 and two SDO-AIA flaring peaks existed for the G1:J3 event. This can also be observed in the zoomed-in \citet{cairns2018}, fig. 7. The radio data had a consistent travel time based on our analysis, being almost identical for the two eruptive events leading to our assertion  that both EUV jet events correlate with the interplanetary bursts. Now, the RHESSI fluxes show a consistent disparity in X-ray energy that is dissipated towards the lower atmosphere. Could this be an indicator that the physical properties of the two events are very dissimilar? We hint that DEM solutions and RHESSI energetic budgets for the two events are of particular interest. A full DEM analysis of the footpoint is discussed separately (see Chap. 3 \citet[][PhD Thesis]{paraschiv2018}, in preparation).  

\subsection{Comparison with recent radio studies}
This work aims to generalize the results of \citet{chifor2008} and \citet{innes2011} who studied recurrent AR jets and their associated Type-III for short time intervals. Recurrent jets from multiple sites were readily associated to Type-III emission by \citet{mulay2016}. Our results build on the supposed hypothesis by offering a large sample correlation of the two solar phenomena. 

 \citet{cairns2018} provide observations of the in-depth X-ray, EUV and higher frequency manifestations of a flaring episode covering a $\sim$15 minute period. In our analysis, we separated the episode into the G1:J2 and G1:J3 events. The authors argued in favor of the standard flare model implying that electron beams are accelerated in or near the reconnection sites but also acknowledged that more evidence is needed to validate the hypothesis. Our analysis adds more weight to the argument by analyzing a  statistically large set of events. The authors acknowledge that special conditions are required to associate radio manifestations with EUV reconnection sites. Our results are optimistic in this regard, where our determination, performed on multiple independent datasets, resulted in correlations for 82\% events, although 9 non-correlated events remain. 
 
 A possible explanation of the special conditions required for generating upwards propagating electron beams is that a homologous flaring site may not always reconnect with an ambient magnetic field that is tied to the outer corona \citep{judge2017}. Event when a coronal connection exist, if the reconnection would occur in denser chromospheric regions, we hypothesize that enhanced collisions might suppress the generation of high energy particles. We are not aware of any report of this happening at our discussed scales. Our discussed Geyser sites are in general associated to X-Ray emission and we assumed that electron acceleration was present (\cref{chap:beam:figcartoon1}). The uncorrelated events may just not be intense enough to produce detectable radio emission in the WIND or STEREO radio dynamic spectra, or were obscured by other radio emission.
 
 Our estimation based on DEM limits of the local plasma places the Geyser's flaring site at a height of maximum 2-3 Mm. This result suggests that the Geyser is located at smaller than previously quoted heights, e.g. \citet{cairns2018} who placed the source at 5-10 Mm. Their lower limit can be considered an upper approximation, but the higher limit is not compatible with a low atmosphere site, and the ``Heights $<$ 10 Mm correspond to the nominal chromosphere'' \citep{cairns2018} does not represent an accurate approximation. We attributed the 2-3 Mm heights to a region in between the high-chromosphere and low corona based on the inferred local conditions. The chromosphere is depicted in hydrostatic numerical experiments (e.g. \citet{fontenla2011} models to be at heights lower than 2 Mm.  Chromospheric structures have been seen extending to heights of $\sim$4 Mm in non polar regions \citep{johannesson1996}. 

 Recent work based on MWA data \citep{mccauley2017} probes into the finer structure of the processes that govern the reconnection site and discuss how a complex magnetic topology can give rise to high frequency beam splitting near or at the coronal source. They derived an average beam travel speeds $v/c=0.2$ using a technique that is independent from the frequency drift rates used in our estimation. Their result is very close to our $v/c=0.16$ estimation. Beam travel speeds in the order of $v/c=0.2$ are typically reported \citep{melendez1999,reid2018}.  We calculated a $\sigma=0.06$ uncertainty  in the beam propagation speed, though this may be skewed by the limited data cadence and event sample. We also remind the reader that our beam travel speed estimation represents a global average across the 41 correlated events. As the propagation speed may be related to the flaring strength, magnetic field strength or topology, it should be evaluated on a case by case basis.  \citet{mccauley2018} used MWA so study three Type-III bursts (electron beams) and fitted them to a high range of propagation speeds, 0.24-0.60 c. The higher limit is well beyond the mildly relativistic regime and proper relativistic effects need to be applied for accurate interpretation. If the higher limit can be proven concrete,  such result is striking, as a typical microflare coronal source does not exhibit adequate non-thermal power for such a high electron beam acceleration. Alternatively, if a stronger reconnective event triggered the burst, the topology may be of particular interest as the conditions for Type-III emission are rather strict.  

\section{Conclusions and Future Prospects}
This study presents data analysis and analytic modeling of a statistically significant sample of recurrent jet microflaring footpoins, designated as Coronal Geysers, that act as sources of upwards accelerated electron beams and interplanetary Type-III radio bursts. A comprehensive database of EUV recurrent jets is presented and temporally correlated with the interplanetary Type-III radio bursts observed by the STEREO-SWAVES and WIND-WAVES instruments. The temporal correlations revealed a systematic time delay of 72$\pm$33 s between the detection of thermal EUV eruptions and the non-thermal generated Type-III radio bursts. We hypothesize that under a standard flare scenario, the electron beams are accelerated at the same time as the EUV jets by recurrent reconnection events at the location of each identified Coronal Geysers.  Using a heliospheric density model adapted to our observations we computed the electron beam propagation times in the inner heliosphere and constrained a modeled time delay to be in the order of 90 s, closely confirming the observations. These results confirmed the scaling of standard flare properties to micro-class flaring, found that recurrent jet-inducing microflare sites are sources of electron acceleration, and demonstrated that `Geysers' are reliable coronal sources of ubiquitous interplanetary Type-III radio bursts.  

Our results represent a global approximation, as they are based on a statistically significant event sample from several different Geyser sites, all rooted near active regions. The accuracy of the EUV to Type-III burst correlation is limited by the instrumental capabilities and constraints in the analytic solution. Improved quality radio data with higher temporal cadence and higher frequency range are needed to probe into the intricate flaring processes of our star. Our future work will focus on closing the gap between higher frequency radio imaging at the source locations and the interplanetary electron beams. As solar energetic events exhibit a high morphological and energetic variability, we believe that the assumed theoretical physical mechanisms can only be scrutinized in solar conditions by utilizing significant datasets. The study of solar (small-scale) flares is an intrinsic topic of solar research due to the close proximity and consequently higher quality observations. As seen, current observational and theoretical results on electron beam generation and propagation show that although a general phenomenological understanding holds true, a unifying mechanism still proves elusive. Further input on particle acceleration, transport and escape into the inner heliosphere in the context of Sun-Earth interconnections is a key aspect that can be extended to more general astrophysics applications.

\acknowledgments
The authors thank the referee for the suggestions which significantly improved this paper.  A.R.P. acknowledges support through Monash University, The Monash School of Mathematical Sciences, the Astronomical Society of Australia and through an Australian Government Research Training Program (RTP) Scholarship. Raw data and calibration instructions are obtained courtesy of NASA/SDO-AIA, STEREO-EUVI and STEREO-SWAVES, and WIND-WAVES science teams. The authors welcome and appreciate the open data policy of the SDO, WIND and STEREO missions. 

\bibliography{bibliography}



\end{document}